\documentclass[aps,prl,twocolumn]{revtex4}
\usepackage{graphicx}
\usepackage{amsmath}
\usepackage{amssymb}
\usepackage{color}

\begin{document}

\title{Superradiance induced particle flow via dynamical gauge coupling}
\author{W. Zheng and N. R. Cooper}
\affiliation{T.C.M. Group, Cavendish Laboratory, J.J. Thomson Avenue, Cambridge CB3 0HE,
United Kingdom }
\date{\today}

\begin{abstract}
We study fermions that are gauge-coupled to a cavity mode via Raman-assisted
hopping in a one dimensional lattice. For an infinite lattice, we find a
superradiant phase with infinitesimal pumping threshold which induces a
directed particle flow. We explore the fate of this flow in a finite lattice
with boundaries, studying the non-equilibrium dynamics including fluctuation
effects. The short time dynamics is dominated by superradiance, while the
long time behaviour is governed by cavity fluctuations. We show that the
steady state in the finite lattice is not unique, and can be understood in
terms of coherent bosonic excitations above a Fermi surface in \textit{real}
space.
\end{abstract}

\maketitle

Quantum matter interacting with gauge fields is a central topic of modern
physics. In cold atom systems, although atoms are charge neutral, Abelian or
non-Abelian synthetic gauge potentials can be simulated by various methods%
\cite{2011revGF}\cite{2013revGF}, such as rotation\cite{2009rotation},
magnetic gradients\cite{2014monopole}, two-photon Raman transitions\cite%
{2009vectorpotential}\cite{2009magneticfield}\cite{2011SOC}, laser-assisted
hopping\cite{Bloch2011}\cite{Bloch2013}\cite{MIT2013}\cite{Bloch2015}, and
lattice \textquotedblleft shaking\textquotedblright \cite{2011Shakinglattice}%
. However, simulation of a dynamical gauge field, possessing its own quantum
dynamics, is still a great challenge\cite{2012DynamicGF}\cite{2013DynamicGF}.

On the other hand, subjecting quantum gases to optical cavities\cite{rev2013}
has drawn a lot of attention in recent years. The coupling between cold
atoms and the quantized cavity modes can dramatically change the properties
of both the atomic gas and the cavity field. For example, a Bose-Einstein
condensate coupled to a cavity can undergo a quantum phase transition to a
supersolid phase. At the same time, the cavity field enters the
\textquotedblleft superradiant" phase with a non-zero expectation value\cite%
{Esslinger2010}\cite{Esslinger2012}\cite{Esslinger2015v2}\cite{Ritsch2002}%
\cite{Domokos2010}. The Bose-Hubbard model inside a cavity exhibits a rich
phase diagram, due to cavity-induced long range interactions between atoms%
\cite{Hemmerich2015r2}\cite{Hemmerich2015}\cite{Esslinger2015}. These
successful experiments have stimulated many theoretical studies in this
direction\cite{2009Simons}\cite{2010Goldbart}\cite{2013Diehl}\cite%
{Simons2014}\cite{Piazza2014}\cite{Zhai2014}\cite{Yi2014}\cite{Pu2014}\cite%
{Yu2015}\cite{Yi2015}\cite{2015hopping}\cite{2016Kollath} \cite{Zhai2015}%
\cite{2016hopping}\cite{2016Donner}. Dissipation of the cavity field,
through photon loss, causes significant back action on the atomic system, on
both its dynamics\cite{Konya2014} and its steady state distribution\cite%
{Griesser2010}\cite{Griesser2011}\cite{Schutz2013}\cite{Piazzal2014}.

In this letter, we study the steady states and the non-equilibrium dynamics
of fermions in a one dimensional cavity-assisted hopping lattice. The phase
of the cavity mode acts on the atoms as a vector potential, which has its
own quantum dynamics controlled by the atom distribution. This system
differs from the models in Refs.\cite{2016Kollath}\cite{2016hopping}, where
the cavity-assisted hopping acts only between two legs of a ladder. Allowing
hopping along an infinite lattice, we find a transition, at infinitesimal
pumping threshold, to a superradiant phase in which the gauge coupling
induces a directed persistent current. In a finite lattice, with open
boundary conditions, we show that there can be no superradiant steady state.
We study the non-equilibrium dynamics in the finite lattice, incorporating
fluctuation effects beyond mean field. On short time scales, particles flow
by coherent hopping as for the infinite lattice, while in the long time
limit, dissipation dominates particle transport and determines the steady
state. Through a mapping to collective bosonic modes in real space, we show
that this steady state is not unique.

\begin{figure}[tbp]
\includegraphics[width=2.8in]
{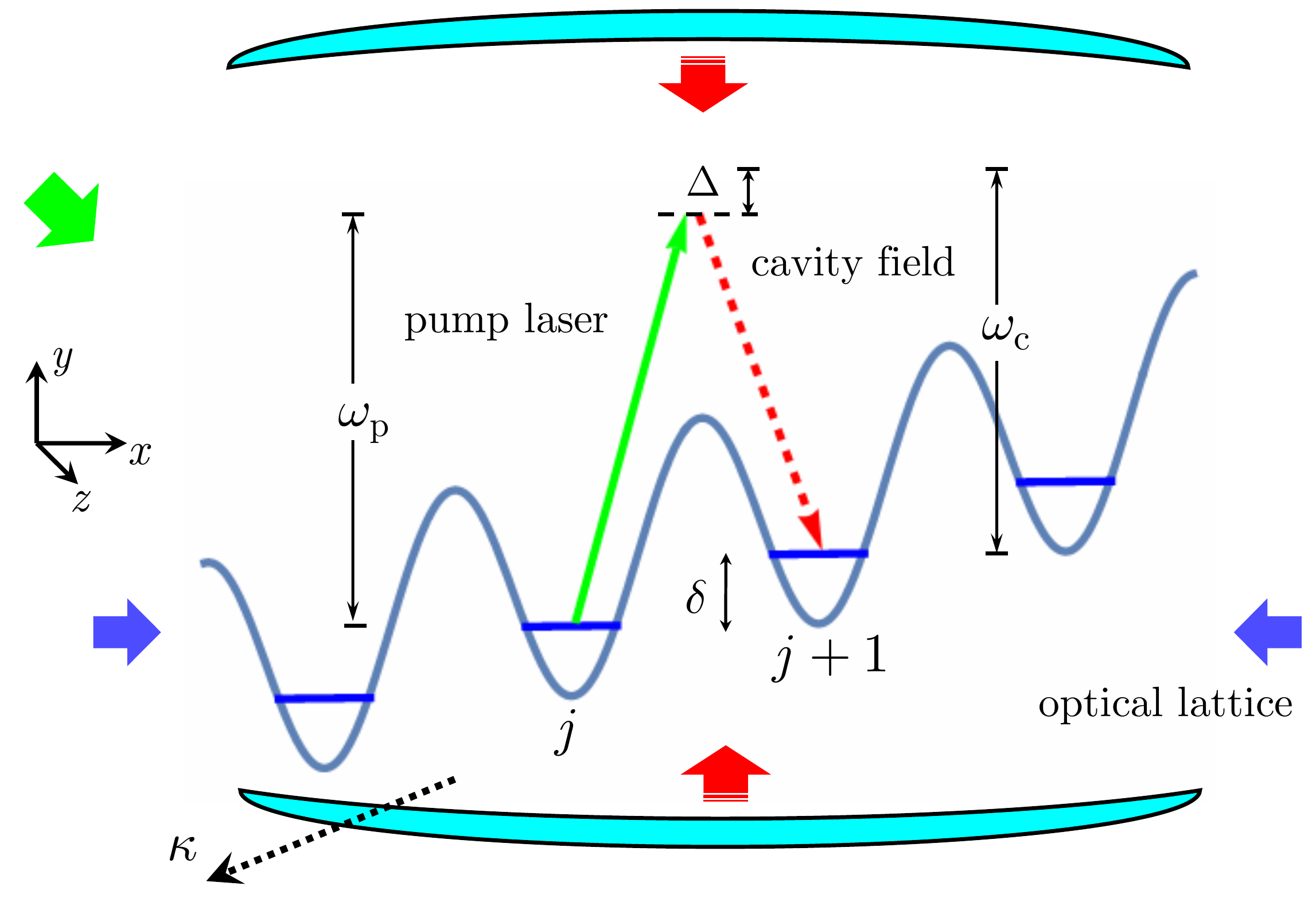}
\caption{The setup for cavity-assisted hopping on a lattice. A large energy
offset $\protect\delta$ prevents direct tunneling between neighbouring
sites. The atoms can hop by a cavity-assisted Raman process, absorbing a
pump photon at $\protect\omega_p$ (solid green arrow) and emitting a photon
at $\protect\omega_p-\protect\delta$ into the cavity (red dashed arrow).
This emission is detuned from the cavity mode, $\protect\omega_c$, by $%
\Delta \ll\protect\delta$. Cavity losses are described by $\protect\kappa$. }
\label{fig1}
\end{figure}

\textit{Model.}-- We consider spinless atoms trapped by an optical lattice
in a high-Q cavity, Fig.\ref{fig1}. The optical lattice is in the $x$%
-direction, while the cavity mode is in the $y$-direction. The atom cloud is
illuminated by a pump laser in the $z$-direction. We consider a strong
transverse confinement to prohibit momentum transfer to the atoms, so the
system is quasi one dimensional. By accelerating the optical lattice or
applying a gradient magnetic field, an energy gradient can be imposed along
the $x$-direction so that direct hopping is suppressed by a large energy
offset $\delta $ between lattice sites. An atom can hop to the right by a
Raman process, absorbing a pump photon ($\omega _{p}$) and emitting at $%
\omega _{p}-\delta $. (We assume $\omega_p$ to be far detuned from the
optical transition so the excited state population can be neglected.) We
consider this emission to be enhanced by a cavity mode tuned close to this
frequency, $\omega _{c}\simeq \omega _{p}-\delta $. (We assume that $\delta $
is sufficiently large that emission at $\omega _{p}+\delta $, corresponding
to a hop to the left, is negligible.) We make a tight-binding approximation
to obtain the effective Hamiltonian ($\hbar =1$ throughout):%
\begin{equation}
\hat{H}=\Delta \hat{a}^{\dag }\hat{a}-\sum_{j=1}^{L-1}\left( \lambda \hat{a}%
^{\dag }\hat{c}_{j+1}^{\dag }\hat{c}_{j}+\lambda ^{\ast }\hat{a}\hat{c}%
_{j}^{\dag }\hat{c}_{j+1}\right) \,.  \label{eq:ham}
\end{equation}%
Here $\hat{a}$ is the field operator of the cavity photon expressed in a
frame rotating at the frequency $\omega _{p}-\delta $ for which intersite
hopping is resonant; $\Delta \equiv \delta -\omega _{p}+\omega _{c}$ is the
detuning of the cavity mode from resonance\footnote{%
The cavity frequency $\omega _{c}=\omega _{c}^{0}+N\epsilon _{c}$ for $N$
atoms includes a frequency shift from that of the empty cavity $\omega
_{c}^{0}$ due to the presence of the atoms. Since $N$ is conserved, this is
just a constant.}; and $\hat{c}_{j}^{(\dag )}$ are fermionic field operators
on lattice sites $j$. (We shall also mention some results for hard-core
bosons.) The cavity-assisted hopping $\lambda \hat{a}^{\dag }$ has a phase
given by the phase difference between the cavity field and the pump laser%
\footnote{%
The pump field oscillates as $e^{i\left( \omega _{p}t+\phi _{p}\right) }$,
while due to the energy gradient, the fermion operator $\hat{c}_{j}$
oscillates as $e^{-ij\delta t}$. These cause the cavity to be driven at an
effective drive frequency $\omega _{p}-\delta $. So the cavity field will
oscillate as $e^{i\left[ \left( \omega _{p}-\delta \right) t+\phi _{c}(t)%
\right] }$. In the rotating frame with this frequency $\omega _{p}-\delta $,
the phase of hopping is $\theta (t)=\phi _{c}(t)-\phi _{p}$.}. We choose to
set the phase of the pump to zero, $\lambda ^{\ast }=\lambda $, such that
the hopping phase equals the phase of the cavity field.\footnote{%
This Hamiltonian can also be realized by the cavity-modulated optical
lattice, similar to \cite{Bloch2011}\cite{Bloch2013}\cite{MIT2013}\cite%
{Bloch2015}. In that case, $\lambda $ has a spatially dependent phase, which
can be eliminated by a local gauge transformation.}

If the cavity were replaced by a second drive laser, at frequency $%
\omega_p-\delta$, such that the cavity field operator is replaced by the
coherent state $\langle \hat{a}\rangle =\alpha =|\alpha |e^{i\theta }$, then
the particles would experience the static Hamiltonian $\hat{H}(\alpha
)=-\lambda \sum_{j}\left( \alpha ^{\ast }\hat{c}_{j+1}^{\dag }\hat{c}
_{j}+\alpha \hat{c}_{j}^{\dag }\hat{c}_{j+1}\right) $. The corresponding
dispersion relation (for an infinite lattice) is
\begin{equation}
E_{k}=-2\lambda \left\vert \alpha \right\vert \cos \left( k+\theta \right)
\label{dispersion}
\end{equation}%
for a particle of momentum $k$. Thus, the phase of the cavity field, $\theta$%
, couples to the particles as a vector potential. In this driven case, the
vector potential is static, set by the phase difference between the two
driving lasers. Henceforth we shall treat the cavity field as dynamical, so
the vector potential inherits its own quantum dynamics, linked to the
distribution of particles. This differs from the cavity-assisted hopping in
Refs.\cite{2015hopping}\cite{2016Kollath}, where the hopping phase is fixed,
and only the amplitude is dynamical.

\textit{Superradiance.}-- The leakage of photons from the cavity requires
the full dynamics to be described by the Lindblad master equation, $\partial
_{t}\rho =-i\left[ \hat{H},\rho \right] +\mathcal{L}[\rho ]$, where $\rho $
is the density matrix, and the Lindblad superoperator reads $\mathcal{L}%
[\rho ]=\kappa \left( 2\hat{a}\rho \hat{a}^{\dag }-\hat{a}^{\dag }\hat{a}%
\rho -\rho \hat{a}^{\dag }\hat{a}\right) $. This describes a cavity photon
loss rate of $2\kappa $. The mean cavity field, $\left\langle \hat{a}%
(t)\right\rangle =\alpha (t)$, evolves as:
\begin{equation}
\partial _{t}\alpha =-i\left( \Delta -i\kappa \right) \alpha +i\lambda K,
\label{cavity_eom}
\end{equation}%
where $K\equiv \langle \hat{K}\rangle $, with $\hat{K}\equiv \sum_{j=1}^{L-1}%
\hat{c}_{j+1}^{\dag }\hat{c}_{j}$ the operator that couples to the cavity
field (\ref{eq:ham}). For steady states, $\partial _{t}\alpha =0$, we obtain
\begin{equation}
\alpha =\frac{\lambda K}{\Delta -i\kappa }.  \label{steady}
\end{equation}%
with $\Delta $, $\kappa $ and $\lambda $ real parameters.

Consider first an infinitely long lattice. In this case, one can write $%
K=\sum_{k}e^{-ik}\left\langle \hat{n}_{k}\right\rangle $, where $\hat{n}_{k}$
counts the number of particles of momentum $k $. These occupations are
conserved, $[\hat{n}_{k},\hat{H}]=0$, so $\lambda K$ in Eq.~(\ref{cavity_eom}%
) can be treated as an external source, determined by the initial momentum
distribution. Provided the initial distribution has $|K|\neq 0$, the steady
state has $|\alpha |\neq 0$, i.e. there is no threshold for superradiance.
This differs from the usual Dicke-type setup\cite{Dicke1954}, where
superradiance appears only above a critical pumping strength.

\begin{figure}[tbp]
\includegraphics[width=3in]
{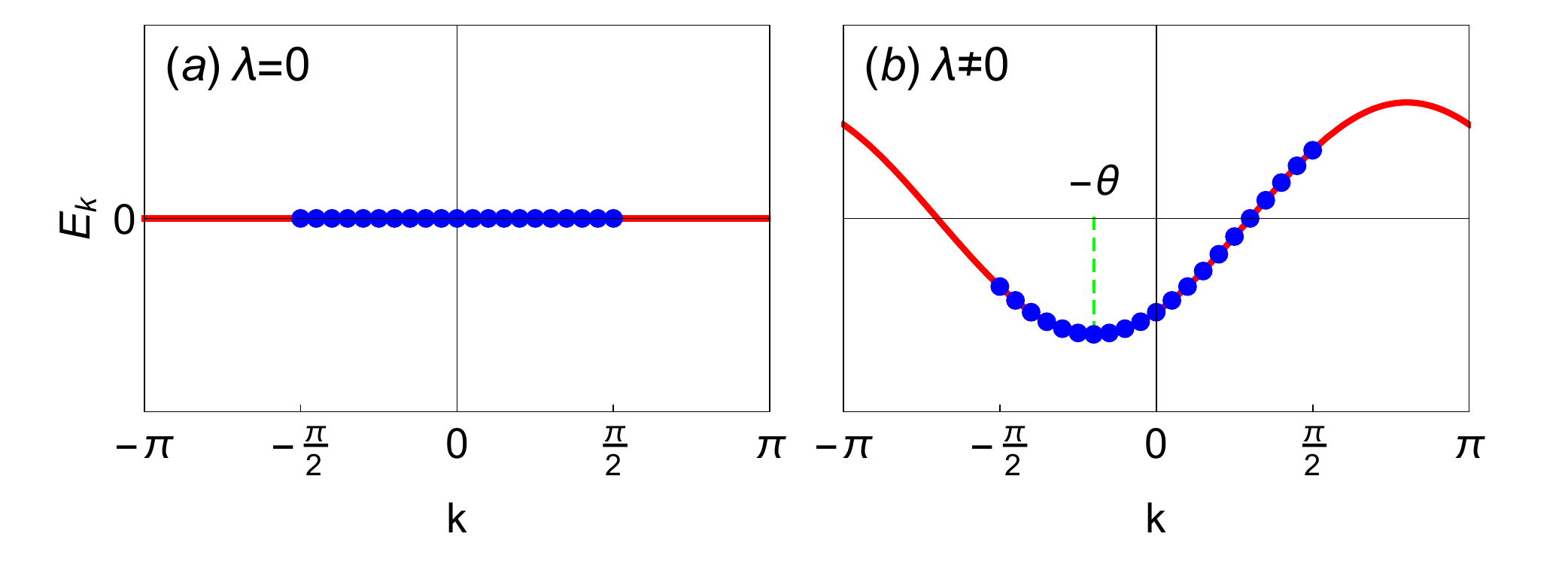}
\caption{Mean field band structure and momentum distribution of fermions.
(a) No pumping field, $\protect\lambda =0$. (b) Non-zero pumping strengh, $%
\protect\lambda \neq 0$. Here $k=-\protect\theta $ is the minimum of the
band.}
\label{fig2}
\end{figure}

Although the cavity field cannot change the momentum distribution of the
atoms, the emergence of superradiance dramatically alters their dispersion (%
\ref{dispersion}). We find that superradiance leads to a directed persistent
current. From (\ref{dispersion}) the particle velocity is $v_{k}=2\lambda
\left\vert \alpha \right\vert \sin \left( k+\theta \right) $, so the total
current, $J=\sum_{k}v_{k}\left\langle \hat{n}_{k}\right\rangle $, may be
written $J=-2\lambda \mathrm{Im}\left( \alpha ^{\ast }K\right) $. Thus,
there will be a non-zero net current if the phases of $K$ and $\alpha $
differ. From Eq.(\ref{steady}), such a phase difference appears whenever
there is cavity loss, $\kappa \neq 0$. For example, consider a half-filled
system with $\left\langle \hat{n}_{k}\right\rangle =\Theta \left( \left\vert
k\right\vert -\pi /2\right) $, see Fig.~\ref{fig2}(a). One finds a real $%
K=L/\pi $, while the phase is $\tan \theta =\kappa /\Delta$. The minimum of
the band is shifted to $k=-\theta $, such that the momentum distribution is
unsymmetrical about it, see Fig.~\ref{fig2}(b). This leads to an imbalance
of left and right moving particles, resulting in a net current to the right.
Thus, the dynamical vector potential self-organizes to induce a particle
current. Indeed, on resonance, $\Delta=0$, the steady state value of cavity
phase $\theta $ maximizes the current ($\frac{\partial J}{\partial \theta }%
=0 $).

The importance of dissipation for the net current can also be seen by
substituting Eq.~(\ref{steady}) into the expression for the total current,
giving $J=2\kappa \left\vert \alpha \right\vert ^{2} $. This has a simple
interpretation. For a cavity occupation of $|\alpha|^2$ the rate of photon
loss is $2\kappa \left\vert \alpha \right\vert ^{2}$. To maintain the
population $|\alpha|^2$, the scattering of pump photons into the cavity
should compensate this loss. Each atom that scatters a photon from pump to
cavity undergoes a hop by one site to the right, thus leading to a net
current of $2\kappa \left\vert \alpha \right\vert ^{2}$.


Now we switch to the finite lattice with open boundary conditions. The
boundaries break translational invariance: momentum is no longer conserved,
so the cavity field can have a feedback on the distribution of the atoms. At
mean field level, the equation-of-motion of the fermionic density matrix, $%
\rho _{ij}(t)=\left\langle \hat{c}_{i}^{\dag }(t)\hat{c}_{j}(t)\right\rangle
$, is
\begin{equation}
\partial _{t}\rho _{ij}(t)=-i\lambda A_{ij}(t),  \label{fermion_eom}
\end{equation}%
where $A_{ij}=\alpha ^{\ast }\rho _{i+1,j}+\alpha \rho _{i-1,j}-\alpha
^{\ast }\rho _{i,j-1}-\alpha \rho _{i,j+1}$. Imposing the boundary
conditions $A_{1,1}=\alpha ^{\ast }\rho _{2,1}-\alpha \rho _{1,2}$ and $%
A_{L,L}=\alpha \rho _{L-1,L}-\alpha ^{\ast }\rho _{L,L-1}$, we can prove $%
\alpha ^{\ast }K=\alpha K^{\ast }$ in any steady state\cite{sup}. Combining
with Eq.(\ref{steady}), we find that the only steady state solution is $%
\alpha =K=0$. The essential physics is that the boundaries preclude the
steady state from carrying a net current, so the net photon scattering rate
from pump mode to cavity mode must vanish. Independent of initial state, the
mean-field steady state is one with $K=0$ and no superradiance, $\alpha =0$.

\begin{figure}[tbp]
\includegraphics[width=3.5in]
{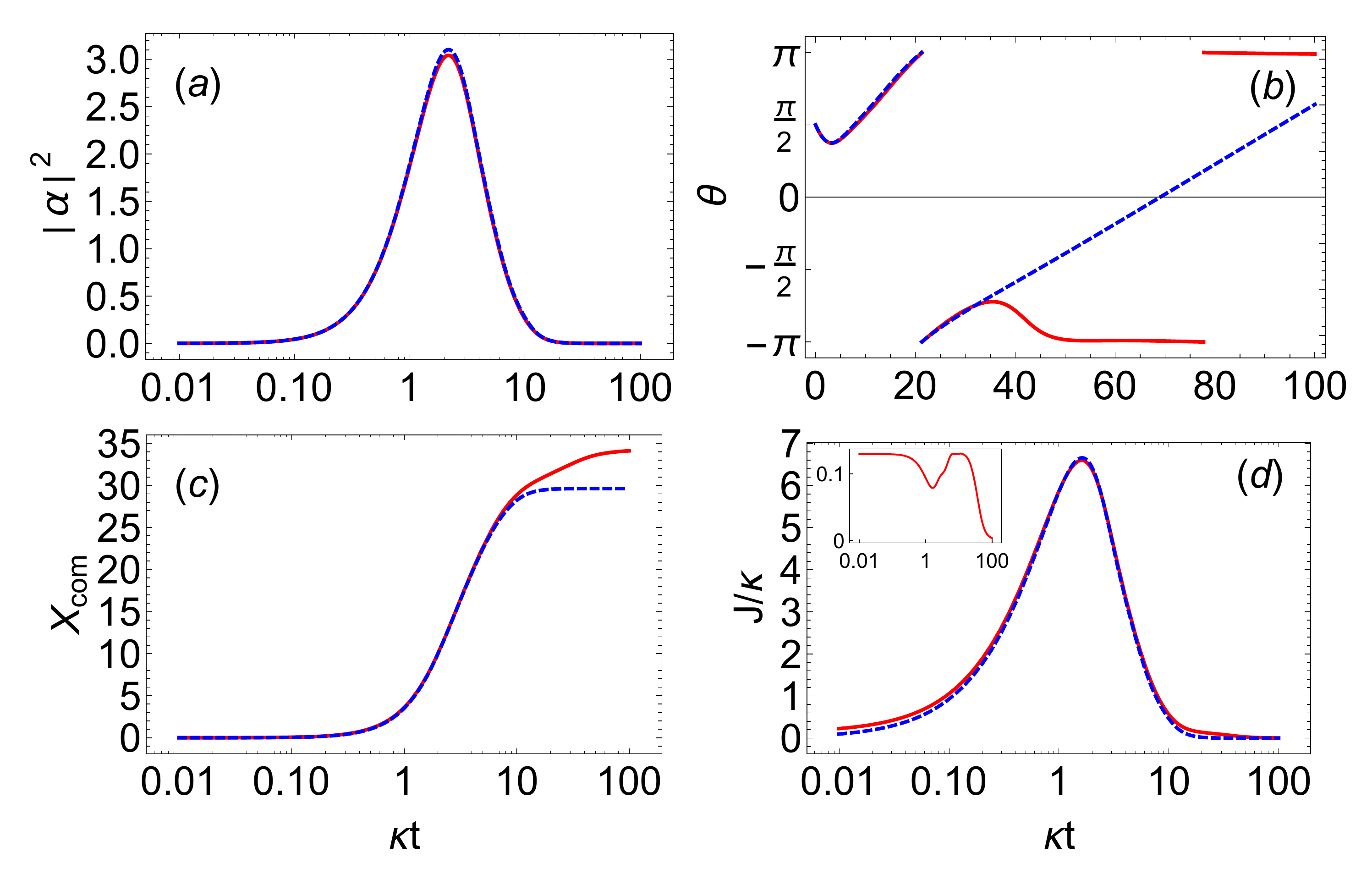}
\caption{The non-equilibrium dynamics of the cavity mode and the fermions.
(a) The cavity field occupation $\left\vert \protect\alpha (t)\right\vert
^{2}$; (b) the phase of $\protect\alpha (t)$; (c) the centre-of-mass of the
fermions; (d) the total current of the fermions. Inset shows the current
beyond mean field, $J^{\mathrm{cl}}+J^{\mathrm{qu}}$. The blue dashed lines
are the mean field results with lattice length $L=20$, particle number $N=5$%
, detuning $\Delta /\protect\kappa =0.5$, and $\protect\lambda /\protect%
\kappa =0.5$. The red solid lines are the results beyond mean field by
solving Eq.(\protect\ref{equ_rou}) with the same parameters.}
\label{fig3}
\end{figure}

\textit{Non-equilibrium dynamics.}-- If the particles start from a state
with non-zero $|K|$, then, even in a finite lattice, the dynamics will first
build up the cavity population $\left\vert \alpha \right\vert ^{2}\neq 0$.
However, any eventual steady state must have $\alpha =K=0$. To understand
how the particles redistribute themselves in a finite lattice, we study the
non-equilibrium dynamics. Combining Eq.(\ref{cavity_eom}) and Eq.(\ref%
{fermion_eom}) describes the coupled mean field dynamics of cavity field and
fermions. However, one can see from Eq.(\ref{fermion_eom}) that for $\alpha
\rightarrow 0$ the fermions cannot hop. This is incorrect, since
fluctuations of the cavity field will cause fermions to hop. To describe
these fluctuation effects in the dynamics, we employ the Keldysh formulation
of open quantum systems\cite{Diehl2015}\cite{book1}, and use the
quasi-particle approximation to obtain the equation-of-motion of the
single-particle density matrix as\cite{sup}:%
\begin{equation}
\partial _{t}\rho _{ij}=-i\lambda A_{ij}(t)+\frac{\kappa \lambda ^{2}}{%
\Delta ^{2}+\kappa ^{2}}B_{ij}(t),  \label{equ_rou}
\end{equation}%
which supplements (\ref{fermion_eom}) with fluctuation corrections, $%
B_{ij}(t)=2\rho _{i-1,j-1}-2\rho _{i,j}-\sum_{l}\left( \rho _{i-1,l-1}\rho
_{l,j}+\rho _{i,l}\rho _{l-1,j-1}\right) +\sum_{l}\left( \rho _{i+1,l+1}\rho
_{l,j}+\rho _{i,l}\rho _{l+1,j+1}\right) $. Here we ignore terms of higher
order than $\lambda^2$, e.g. cavity-induced interactions between particles
at order $\lambda ^{4}$, which is valid for $\lambda \ll \kappa $. From the
diagonal elements, i.e. the particle density $\rho _{i}=\rho _{ii}$, and the
continuity equation, $\partial _{t}\rho _{i}+J_{i}-J_{i-1}=0$, we derive the
current $J_{i}$. We find that the current can be separated into three parts,
$J_{i}=J_{i}^{\mathrm{sr}}+J_{i}^{\mathrm{cl}}+J_{i}^{\mathrm{qu}}$, where $%
J_{i}^{\mathrm{sr}}=-\lambda \mathrm{Im}\left( \alpha ^{\ast }\rho
_{i+1,i}\right) $, is the superradiant current as in mean field. The current
$J_{i}^{\mathrm{cl}}=\frac{2\kappa \lambda ^{2}}{\Delta ^{2}+\kappa ^{2}}%
\left( 1-\rho _{i+1}\right) \rho _{i}$ describes the semiclassical current,
subject to Pauli blocking, arising from dissipative losses of the
fluctuating cavity mode. This current is precisely that for classical
driven-dissipative models such as the asymmetric exclusion process (ASEP)%
\cite{ASEP2006}\cite{ASEP2007}, which has interesting dynamical phase
transitions sensitive to boundary conditions. The contribution $J_{i}^{%
\mathrm{qu}}=-\frac{2\kappa \lambda ^{2}}{\Delta ^{2}+\kappa ^{2}}%
\sum_{l\neq i}\mathrm{Re}\left( \rho _{i+1,l+1}\rho _{l,i}\right) $, is a
quantum correction to the semiclassical current induced by correlations and
involving long-range coherence imposed by the fact that the cavity mode
couples to all atoms. Here we see that even when the superradiance vanishes,
$\alpha =0$, the fluctuations of the cavity mode can induce a nonzero
current $J_{i}^{\mathrm{cl}}+J_{i}^{\mathrm{qu}}$.

We have solved Eq.(\ref{equ_rou}) combined with Eq.(\ref{cavity_eom})
numerically. Representative results are plotted in Fig.\ref{fig3}. We choose
the initial state to be the groundstate of $N$ free fermions in a finite
lattice with non-zero hopping, and the cavity mode empty. Because this
initial particle state has coherence in real space, $K$ is non-zero and,
according to Eq.(\ref{cavity_eom}), it will first generate a superradiant
state. Indeed, we find that the cavity occupation $|\alpha (t)|^{2}$ grows
from zero, and reaches its maximum during a time interval $\tau_{1}\sim
\kappa ^{-1}$ Fig.\ref{fig3}(a). After that, due to the cavity loss, the
superradiance decays to zero on a time scale $\tau_{2}\sim \frac{\kappa \pi
^{2}\nu \left( 1-\nu \right) }{2\lambda ^{2}\sin ^{2}\left( \pi \nu \right) }
$, where $\nu $ is the particle filling. This superradiant pulse is similar
to those observed by illuminating degenerate quantum gases in free space\cite%
{1999Ketterle}\cite{2011Zhang}. Thus, the dynamics can be separated into two
regimes. At short times $t\lesssim \tau_{1}+\tau_{2}$, the particle dynamics
is dominated by coherent hopping, which is assisted by the mean field part
of the cavity field. In this regime, the centre-of-mass of the fermions, $X_{%
\mathrm{com}}=\sum_{j}j\left\langle \hat{c}_{j}^{\dag }\hat{c}%
_{j}\right\rangle $, increases quickly, see Fig.\ref{fig3}(c). At long
times, $t\gg \tau_{1}+\tau_{2}$, the superradiance dies, and the particle
dynamics is governed by the dissipative hopping. See Fig.\ref{fig3}(c),
after the collapse of superradiance, the mean field solution of $X_{\mathrm{%
com}}(t)$ saturates, while the solution including fluctuations shows that $%
X_{\mathrm{com}}(t)$ still grows slowly, until finally reaching a slightly
larger final steady-state value.

\begin{figure}[tbp]
\includegraphics[width=3in]
{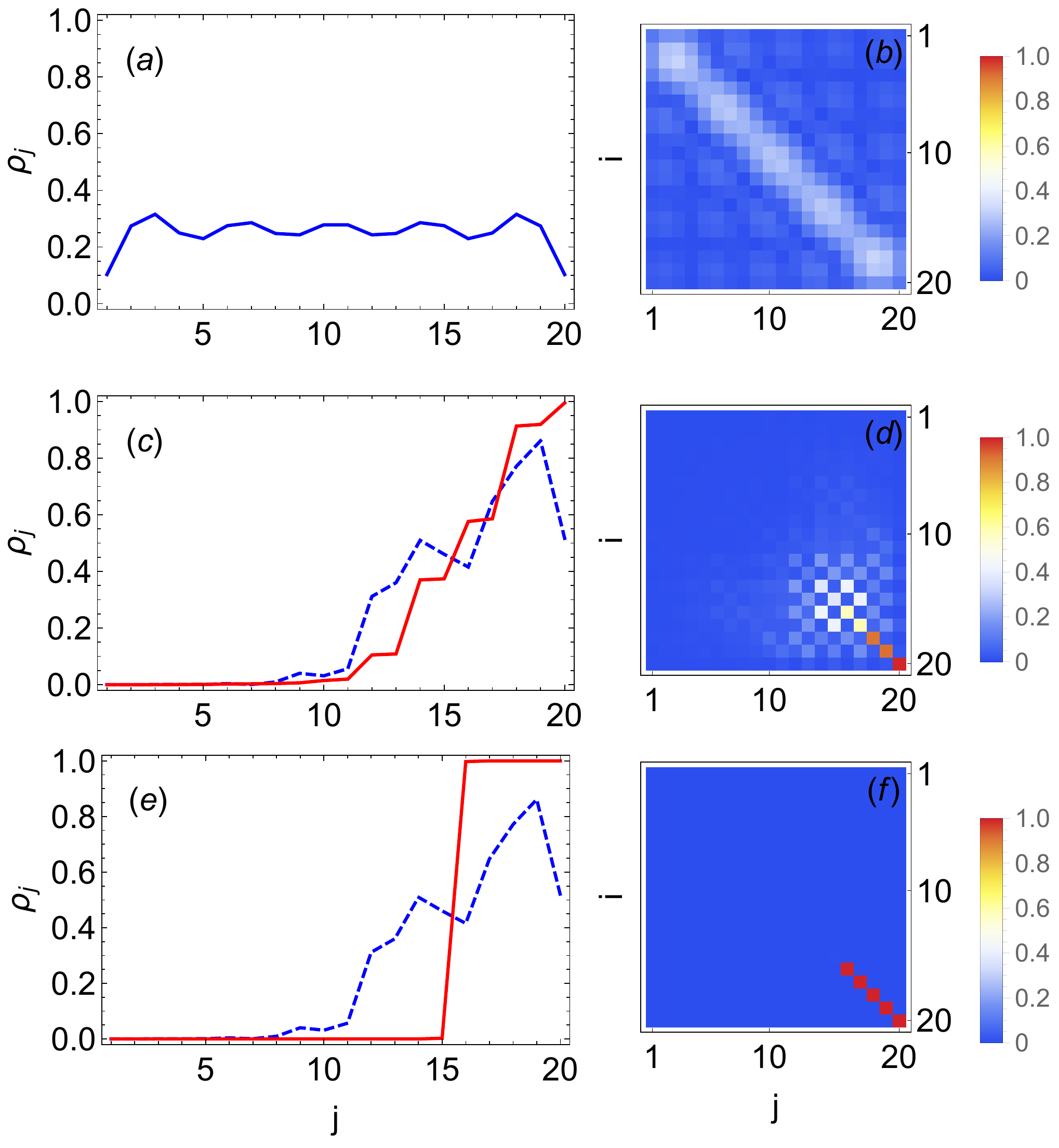}
\caption{(a) The density distribution of initial state. (b) The absolute
value of single-particle density matrix of initial state, $\left\vert
\protect\rho _{ij}(0)\right\vert $. (c)(d) The final density distribution
and density matrix of fermions (beyond mean field) after a long time
evolution $\protect\kappa t=1000$. (e)(f) The final density distribution and
density matrix of hard core boson starting from same initial state. Blue
dashed lines are the mean field results, while the red solid lines are the
solutions beyond mean field. The lattice length $L=20$, particle number $N=5$%
, detuning $\Delta/\protect\kappa =0.5$, and $\protect\lambda /\protect%
\kappa =0.5$.}
\label{fig4}
\end{figure}

To explore the final steady states that are reached long after superradiance
has vanished (so $J^{\mathrm{sr}}=0$), in Fig.\ref{fig4}(c,d) we plot the
density distribution and the density matrix for the fermions at very late
times, $\kappa t=1000$. Before discussing the results for fermions, it is
helpful to consider the final steady state for the case of hard core bosons,
see Fig.\ref{fig4}(e,f). For hard core bosons, the steady-state density
distribution is a simple step function, i.e. the rightmost $N$ sites are
fully populated, while others have no population. The single-particle
density matrix shows no coherence. Since all the particles are blocked at
the right side, no particle can hop to the right due to the hard core
repulsion, giving zero semiclassical current, $J^{\mathrm{cl}}=0$. Vanishing
coherence indicates the quantum correlation current $J^{\mathrm{qu}}$ is
also zero. The situation is different in the case of fermions. As seen in
Fig.\ref{fig4}(c,d), the density distribution is not a step function, while
the density matrix retains non-zero correlations. In this case, the
semiclassical current and its quantum correction do not separately vanish,
instead, they cancel each other in the steady state, with $J^{\mathrm{cl}%
}+J^{\mathrm{qu}}=0$. The quantum correction current and semiclassical
current counteract each other in the case of fermions, while they add
together in the case of bosons.

\textit{Steady states.}-- To understand the steady states, we adiabatically
eliminate the cavity field\cite{Ritsch2005}, making use of the fact that
superradiance is absent, $\alpha =\langle \hat{a}\rangle =0$. We obtain the
master equation for the fermion density matrix, $\partial _{t}\rho _{\mathrm{%
f}}=-i\left[ \hat{H}_{\mathrm{eff}},\rho _{\mathrm{f}}\right] +\frac{\kappa
\lambda ^{2}}{\Delta ^{2}+\kappa ^{2}}\left( 2\hat{L}_{\mathrm{eff}}\rho _{%
\mathrm{f}}\hat{L}_{\mathrm{eff}}^{\dag }-\hat{L}_{\mathrm{eff}}^{\dag }\hat{%
L}_{\mathrm{eff}}\rho _{\mathrm{f}}-\rho _{\mathrm{f}}\hat{L}_{\mathrm{eff}%
}^{\dag }\hat{L}_{\mathrm{eff}}\right) $, where $\hat{L}_{\mathrm{eff}}=\hat{%
K}$ is the effective Liouvillian operator, and the effective Hamiltonian
reads $\hat{H}_{\mathrm{eff}}=-\frac{\kappa \lambda ^{2}}{\Delta ^{2}+\kappa
^{2}}\hat{K}^{\dag }\hat{K}$. Any pure state $\left\vert D\right\rangle $
for which $\hat{H}_{\mathrm{eff}}\left\vert D\right\rangle =E\left\vert
D\right\rangle $ and $\hat{L}_{\mathrm{eff}}\left\vert D\right\rangle =0$ is
a steady state\cite{Diehl2008}\cite{Zoller2010}. Here, these two conditions
reduce to $\hat{K}\left\vert D\right\rangle =0$. It can readily be verified
that the step function state, $\left\vert \mbox{step}\right\rangle \equiv
\prod_{j=1}^{N}\hat{c}_{L-N+j}^{\dag }\left\vert 0\right\rangle $, in which
the $N$ particles occupy the $N$ states furthest to the right, is a steady
state. However we can also find other steady states. To construct these we
define the bosonic operators%
\begin{equation}
\hat{b}_{s}^{\dag }=\sum_{j=s+1}^{L}\hat{c}_{j-s}^{\dag }\hat{c}_{j},
\end{equation}%
where $s=1,\cdots ,L-1$. These are analogous to the bosonic operators used
to solve the Tomonaga-Luttinger model, but now with real space separation $s$
replacing momentum and with $|\mbox{step}\rangle $ viewed as a Fermi sea in
\textit{real space} (with states occupied for $j>L-N$ and empty for $j\leq
L-N$). One can verify that the state with one particle-hole excitation above
this Fermi sea, created by applying one bosonic operator $\hat{b}_{s}^{\dag
}\left\vert \mbox{step}\right\rangle $, is a steady state, via $\hat{K}\hat{b%
}_{s}^{\dag }\left\vert \mbox{step}\right\rangle =\left[ \hat{K},\hat{b}%
_{s}^{\dag }\right] \left\vert \mbox{step}\right\rangle =0$ if $s\neq 1$ and
$s\leq \mathrm{Min}\left( L-N,N\right) $. Similarly, for $n_{\mathrm{b}}$
such bosonic excitations the states $\prod_{\alpha =1}^{n_{\mathrm{b}}}\hat{b%
}_{s_{\alpha }}^{\dag }\left\vert \mbox{step}\right\rangle $, are steady
states provided $s_{\alpha }\neq 1$, and $\sum_{\alpha }s_{\alpha }\leq
\mathrm{Min}\left( L-N,N\right) $. Thus, we can construct a large number of
steady states. For $N,(L-N)\rightarrow \infty $ (when all relevant states
can be described by bosonic modes), any state that does not involve
occupation of the $s=1$ boson is a steady state. This arises because the
Liouvillian operator $\hat{K}$ equals the bosonic annihilation operator $%
\hat{b}_{1}$, so dissipation can only damp the $s=1$ collective mode. This
differs from models involving coupling to a macroscopic number of
dissipation channels\cite{Diehl2008}\cite{Zoller2010}, which lead to unique
steady states. The large number of steady states for our model means that
different initial conditions will lead to different final steady states.

\textit{Final remarks.}-- Our proposal explores one natural route to a
synthetic dynamic gauge coupling in cold atom systems, using elements that
can be realized in current experimental conditions. The dynamics of the
superradiance can be observed by detecting the photons leaving the cavity,
while the redistribution of the fermions could be measured by recently
developed fermionic in-situ imaging in optical lattices\cite{Insitu01}\cite%
{Insitu02}\cite{Insitu03}\cite{Insitu04}\cite{Insitu05}\cite{Zwierlein2016}%
\textbf{.} It is natural to generalize the setup to the two dimensional
case, where the dynamic vector potential can be made spatially dependent to
realize a dynamic magnetic field.

\acknowledgments{We are grateful to Andreas Nunnenkamp for helpful discussions and comments.
This work was supported by ESPRC Grant No. EP/K030094/1.}

\begin{widetext}

\section{ Supplemental material }

\subsection{The fate of superradiance in a finite lattice with open boundary
conditions}

From the master equation, we obtain the equations-of-motion of all operators
as
\begin{eqnarray}
i\partial _{t}\hat{a} &=&\left( \Delta -i\kappa \right) \hat{a}-\lambda
\sum_{j=1}^{L-1}\hat{c}_{j+1}^{\dag }\hat{c}_{j},  \label{eom_cavity} \\
i\partial _{t}\hat{c}_{j} &=&-\lambda \hat{a}^{\dag }\hat{c}_{j-1}-\lambda
\hat{a}\hat{c}_{j+1},\text{\ }\left( j\neq 1,L\right) .  \label{eom_fermion}
\end{eqnarray}%
The open boundary conditions are encoded in the equations-of-motion of $\hat{%
c}_{1}$ and $\hat{c}_{L}$,%
\begin{eqnarray}
i\partial _{t}\hat{c}_{1} &=&-\lambda \hat{a}\hat{c}_{2}, \\
i\partial _{t}\hat{c}_{L} &=&-\lambda \hat{a}^{\dag }\hat{c}_{L-1}.
\end{eqnarray}%
Using these equations-of-motion, one obtains the evolution of local fermion
density as%
\begin{eqnarray}
\partial _{t}\left\langle \hat{c}_{j}^{\dag }\hat{c}_{j}\right\rangle
&=&i\lambda \left\langle \hat{a}^{\dag }\right\rangle \left( \left\langle
\hat{c}_{j}^{\dag }\hat{c}_{j-1}\right\rangle -\left\langle \hat{c}%
_{j+1}^{\dag }\hat{c}_{j}\right\rangle \right) +\lambda \left\langle \hat{a}%
\right\rangle \left( \left\langle \hat{c}_{j}^{\dag }\hat{c}%
_{j+1}\right\rangle -\left\langle \hat{c}_{j-1}^{\dag }\hat{c}%
_{j}\right\rangle \right) , \\
\partial _{t}\left\langle \hat{c}_{1}^{\dag }\hat{c}_{1}\right\rangle
&=&-i\lambda \left\langle \hat{a}^{\dag }\right\rangle \left\langle \hat{c}%
_{2}^{\dag }\hat{c}_{1}\right\rangle +i\lambda \left\langle \hat{a}%
\right\rangle \left\langle \hat{c}_{1}^{\dag }\hat{c}_{2}\right\rangle , \\
\partial _{t}\left\langle \hat{c}_{L}^{\dag }\hat{c}_{L}\right\rangle
&=&i\lambda \left\langle \hat{a}^{\dag }\right\rangle \left( \left\langle
\hat{c}_{L}^{\dag }\hat{c}_{L-1}\right\rangle \right) -i\lambda \left\langle
\hat{a}\right\rangle \left\langle \hat{c}_{L-1}^{\dag }\hat{c}%
_{L}\right\rangle .
\end{eqnarray}%
In the steady state, $\partial _{t}\left\langle \hat{c}_{j}^{\dag }\hat{c}%
_{j}\right\rangle =0$, then we have
\begin{eqnarray}
\alpha ^{\ast }\left( \left\langle \hat{c}_{j+1}^{\dag }\hat{c}%
_{j}\right\rangle -\left\langle \hat{c}_{j}^{\dag }\hat{c}%
_{j-1}\right\rangle \right) &=&\alpha \left( \left\langle \hat{c}_{j}^{\dag }%
\hat{c}_{j+1}\right\rangle -\left\langle \hat{c}_{j-1}^{\dag }\hat{c}%
_{j}\right\rangle \right) , \\
\alpha ^{\ast }\left\langle \hat{c}_{2}^{\dag }\hat{c}_{1}\right\rangle
&=&\alpha \left\langle \hat{c}_{1}^{\dag }\hat{c}_{2}\right\rangle , \\
\alpha ^{\ast }\left\langle \hat{c}_{L}^{\dag }\hat{c}_{L-1}\right\rangle
&=&\alpha \left\langle \hat{c}_{L-1}^{\dag }\hat{c}_{L}\right\rangle .
\end{eqnarray}%
From these equations, it is straightforward to see $\alpha ^{\ast
}\left\langle \hat{c}_{j+1}^{\dag }\hat{c}_{j}\right\rangle =\alpha
\left\langle \hat{c}_{j}^{\dag }\hat{c}_{j+1}\right\rangle $, where $%
j=1,\cdots ,L-1$. That gives%
\begin{equation}
\alpha ^{\ast }\sum_{j=1}^{L-1}\left\langle \hat{c}_{j+1}^{\dag }\hat{c}%
_{j}\right\rangle =\alpha \sum_{j=1}^{L-1}\left\langle \hat{c}_{j}^{\dag }%
\hat{c}_{j+1}\right\rangle .
\end{equation}%
This is nothing but $\alpha ^{\ast }K=\alpha K^{\ast }$. That indicates the
phase shift between $\alpha $ and $K$ is either $0$ or $\pi $. So there can
be no superradiant steady state in a finite lattice with open boundary
conditions.

\subsection{Quantum kinetic equation of the single-particle density matrix of
the fermions}

From the equations-of-motion (\ref{eom_cavity}) and (\ref{eom_fermion}), we
obtain%
\begin{eqnarray}
i\partial _{t}\left\langle \hat{a}\right\rangle  &=&\left( \Delta -i\kappa
\right) \left\langle \hat{a}\right\rangle -\lambda \sum_{j}\left\langle \hat{%
c}_{j+1}^{\dag }\hat{c}_{j}\right\rangle ,  \label{eom_alpha} \\
i\partial _{t}\left\langle \hat{c}_{i}^{\dag }\hat{c}_{j}\right\rangle
&=&\lambda \left[ \left\langle \hat{a}^{\dag }\hat{c}_{i+1}^{\dag }\hat{c}%
_{j}\right\rangle +\left\langle \hat{a}\hat{c}_{i-1}^{\dag }\hat{c}%
_{j}\right\rangle -\left\langle \hat{a}^{\dag }\hat{c}_{i}^{\dag }\hat{c}%
_{j-1}\right\rangle -\left\langle \hat{a}\hat{c}_{i}^{\dag }\hat{c}%
_{j+1}\right\rangle \right] .  \label{eom_dm}
\end{eqnarray}%
Then we separate the mean field and the fluctuation parts of the cavity
field as%
\begin{equation}
\hat{a}(t)=\left\langle \hat{a}(t)\right\rangle +\delta \hat{a}(t)=\alpha
(t)+\delta \hat{a}(t),
\end{equation}%
where the fluctuation operator satisfies the usual bosonic commutation
relation, $\left[ \delta \hat{a},\delta \hat{a}^{\dag }\right] =1$. Then Eq.
(\ref{eom_alpha}) and (\ref{eom_dm}) can be rewritten into
\begin{eqnarray}
i\partial _{t}\alpha  &=&\left( \Delta -i\kappa \right) \alpha -\lambda
\sum_{j}\rho _{j+1,j},  \label{QKE_cavity} \\
i\partial _{t}\rho _{i,j} &=&\lambda \left( \alpha ^{\ast }\rho
_{i+1,j}+\alpha \rho _{i-1,j}-\alpha ^{\ast }\rho _{i,j-1}-\alpha \rho
_{i,j+1}\right)   \notag \\
&&+\lambda \left[ \left\langle \delta \hat{a}^{\dag }\hat{c}_{i+1}^{\dag }%
\hat{c}_{j}\right\rangle +\left\langle \delta \hat{a}\hat{c}_{i-1}^{\dag }%
\hat{c}_{j}\right\rangle -\left\langle \delta \hat{a}^{\dag }\hat{c}%
_{i}^{\dag }\hat{c}_{j-1}\right\rangle -\left\langle \delta \hat{a}\hat{c}%
_{i}^{\dag }\hat{c}_{j+1}\right\rangle \right]
\end{eqnarray}%
where $\rho _{i,j}(t)=\left\langle \hat{c}_{i}^{\dag }(t)\hat{c}%
_{j}(t)\right\rangle $, is the single-particle density matrix of fermions.
Here the mean field part of the cavity field behaves as an time-dependent
potential, which acts on the single-particle dynamics of the fermions.
However, these equations are not closed. To deal with this problem, we
introduce the Keldysh Green's function of fermions, which is defined as%
\begin{equation}
iG_{ij}^{K}\left( t_{2},t_{1}\right) =\left\langle \left[ \hat{c}_{i}(t_{2}),%
\hat{c}_{j}^{\dag }(t_{1})\right] \right\rangle .
\end{equation}%
The single-particle density matrix can be calculated from this Keldysh Green
function by%
\begin{eqnarray}
\rho _{ij}(t) &=&\frac{1}{2}\left[ \delta _{ji}-iG_{ji}^{K}\left( t,t\right) %
\right] ,  \notag \\
&=&\frac{1}{2}\left[ \delta _{ji}-i\int \frac{d\omega }{2\pi }%
G_{ji}^{K}\left( t,\omega \right) \right] ,
\end{eqnarray}%
where $G_{ji}^{K}\left( t,\omega \right) $ is the Wigner transformation of $%
G_{ji}^{K}\left( t_{2},t_{1}\right) $, which is defined as $G_{ji}^{K}\left(
t,\omega \right) =\int d\tau e^{i\omega \tau }G_{ji}^{K}\left( t+\tau
/2,t-\tau /2\right) $. So once having the quantum kinetic equation of the
Keldysh Green's function, we can immediately obtain the evolution of the
single-particle density matrix. The quantum kinetic equation in terms of the
Keldysh Green's function is given by\cite{2007Rammer}%
\begin{equation}
\left[ G_{0}^{-1}-\mathrm{Re}\Sigma ^{R},G^{K}\right] _{\circ }+\left[
\mathrm{Re}G^{R},\Sigma ^{K}\right] _{\circ }=\frac{1}{2}i\left\{ A,\Sigma
^{K}\right\} _{\circ }-\frac{1}{2}i\left\{ \Gamma ,G^{K}\right\} _{\circ },
\end{equation}%
where $G_{0}^{-1}$ is the free Green function, $G^{R\left( A\right) }$ is
the retarded(advanced) Green's function, defined by%
\begin{eqnarray}
G_{ij}^{R}\left( t_{2},t_{1}\right)  &=&-i\Theta \left( t_{2}-t_{1}\right)
\left\langle \left\{ \hat{c}_{i}(t_{2}),\hat{c}_{j}^{\dag }(t_{2})\right\}
\right\rangle , \\
G_{ij}^{A}\left( t_{2},t_{1}\right)  &=&-i\Theta \left( t_{1}-t_{2}\right)
\left\{ \hat{c}_{i}(t_{2}),\hat{c}_{j}^{\dag }(t_{2})\right\} ,
\end{eqnarray}%
and $\Sigma ^{K\left( R,A\right) }$ is the Keldysh(retarded, advanced)
self-energy. The spectrum function $A\left( t_{2},t_{1}\right) $ and
lifetime function $\Gamma \left( t_{2},t_{1}\right) $ are given by
\begin{eqnarray}
\Gamma \left( t_{2},t_{1}\right)  &=&i\left[ \Sigma ^{R}\left(
t_{2},t_{1}\right) -\Sigma ^{A}\left( t_{2},t_{1}\right) \right] , \\
A\left( t_{2},t_{1}\right)  &=&i\left[ G^{R}\left( t_{2},t_{1}\right)
-G^{A}\left( t_{2},t_{1}\right) \right] ,
\end{eqnarray}%
The commutator(anti-commutator) is defined as $\left[ f_{1},f_{2}\right]
_{\circ }=f_{1}\circ f_{2}-f_{2}\circ f_{1}$($\left\{ f_{1},f_{2}\right\}
_{\circ }=f_{1}\circ f_{2}+f_{2}\circ f_{1}$), where $f_{1}\circ f_{2}$
denotes the time-space convolution of the two-point function. Employing the
quasi-particle approximation, i.e. ignoring the self-energy term in the left
hand side, we have:%
\begin{equation}
\left[ G_{0}^{-1},G^{K}\right] _{\circ }=\frac{1}{2}i\left\{ A,\Sigma
^{K}\right\} _{\circ }-\frac{1}{2}i\left\{ \Gamma ,G^{K}\right\} _{\circ },
\end{equation}%
The left hand side represents the drift of quasi-particles, while the right
hand side represents the collision integral. Making the Wigner
transformation, and using the gradient approximation, we obtain:%
\begin{eqnarray}
i\partial _{t}G_{j_{2}j_{1}}^{K}\left( t,\omega \right)  &=&-\sum_{l}\left\{
h_{j_{2}l}(t)G_{lj_{1}}^{K}\left( t,\omega \right) -G_{j_{2}l}^{K}\left(
t,\omega \right) h_{lj_{1}}(t)\right\}   \notag \\
&&+\frac{1}{2}i\sum_{l}\left[ A_{j_{2}l}(t,\omega )\Sigma
_{lj_{1}}^{K}\left( t,\omega \right) +\Sigma _{j_{2}l}^{K}\left( t,\omega
\right) A_{lj_{1}}(t,\omega )\right]   \notag \\
&&-\frac{1}{2}i\sum_{l}\left[ \Gamma _{j_{2}l}(t,\omega
)G_{lj_{1}}^{K}\left( t,\omega \right) +G_{j_{2}l}^{K}\left( t,\omega
\right) \Gamma _{lj_{1}}(t,\omega )\right] .  \label{QKE_GK}
\end{eqnarray}

To go further, we have to calculate the Keldysh self-energy $\Sigma
_{ij}^{K}\left( t,\omega \right) $ and lifetime function $\Gamma
_{ij}(t,\omega )$. By ignoring the correction of the vertex function, we can
express the self-energy of fermions as\cite{2010Altland}%
\begin{eqnarray}
\Sigma _{j_{2}j_{1}}^{R}\left( t_{2},t_{1}\right)  &=&i\frac{1}{2}\lambda
^{2}\left[ D^{K}\left( t_{2},t_{1}\right) G_{j_{2}+1,j_{1}+1}^{R}\left(
t_{2},t_{1}\right) +D^{K}\left( t_{1},t_{2}\right)
G_{j_{2}-1,j_{1}-1}^{R}\left( t_{2},t_{1}\right) \right]   \notag \\
&&+i\frac{1}{2}\lambda ^{2}\left[ D^{R}\left( t_{2},t_{1}\right)
G_{j_{2}+1,j_{1}+1}^{K}\left( t_{2},t_{1}\right) +D^{A}\left(
t_{1},t_{2}\right) G_{j_{2}-1,j_{1}-1}^{K}\left( t_{2},t_{1}\right) \right] ,
\label{se_f_r} \\
\Sigma _{j_{2}j_{1}}^{A}\left( t_{2},t_{1}\right)  &=&i\frac{1}{2}\lambda
^{2}\left[ D^{K}\left( t_{2},t_{1}\right) G_{j_{2}+1,j_{1}+1}^{A}\left(
t_{2},t_{1}\right) +D^{K}\left( t_{1},t_{2}\right)
G_{j_{2}-1,j_{1}-1}^{A}\left( t_{2},t_{1}\right) \right]   \notag \\
&&+i\frac{1}{2}\lambda ^{2}\left[ D^{A}\left( t_{2},t_{1}\right)
G_{j_{2}+1,j_{1}+1}^{K}\left( t_{2},t_{1}\right) +D^{R}\left(
t_{1},t_{2}\right) G_{j_{2}-1,j_{1}-1}^{K}\left( t_{2},t_{1}\right) \right] ,
\label{se_f_a} \\
\Sigma _{j_{2}j_{1}}^{K}\left( t_{2},t_{1}\right)  &=&i\frac{1}{2}\lambda
^{2}\left[ D^{K}\left( t_{2},t_{1}\right) G_{j_{2}+1,j_{1}+1}^{K}\left(
t_{2},t_{1}\right) +D^{K}\left( t_{1},t_{2}\right)
G_{j_{2}-1,j_{1}-1}^{K}\left( t_{2},t_{1}\right) \right]   \notag \\
&&+i\frac{1}{2}\lambda ^{2}\left[ D^{R}\left( t_{2},t_{1}\right)
G_{j_{2}+1,j_{1}+1}^{R}\left( t_{2},t_{1}\right) +D^{A}\left(
t_{1},t_{2}\right) G_{j_{2}-1,j_{1}-1}^{R}\left( t_{2},t_{1}\right) \right]
\notag \\
&&+i\frac{1}{2}\lambda ^{2}\left[ D^{A}\left( t_{2},t_{1}\right)
G_{j_{2}+1,j_{1}+1}^{A}\left( t_{2},t_{1}\right) +D^{R}\left(
t_{1},t_{2}\right) G_{j_{2}-1,j_{1}-1}^{A}\left( t_{2},t_{1}\right) \right] ,
\label{se_f_k}
\end{eqnarray}%
Here $D^{K\left( R,A\right) }\left( t_{2},t_{1}\right) $ is the full Green's
function of the cavity fluctuation, which is defined as%
\begin{eqnarray}
D^{R}\left( t_{2},t_{1}\right)  &=&-i\Theta \left( t_{2}-t_{1}\right)
\left\langle \left[ \delta \hat{a}(t_{2}),\delta \hat{a}^{\dag }(t_{1})%
\right] \right\rangle , \\
D^{A}\left( t_{2},t_{1}\right)  &=&-i\Theta \left( t_{1}-t_{2}\right)
\left\langle \left[ \delta \hat{a}(t_{2}),\delta \hat{a}^{\dag }(t_{1})%
\right] \right\rangle , \\
D^{K}\left( t_{2},t_{1}\right)  &=&-i\left\langle \left\{ \delta \hat{a}%
(t_{2}),\delta \hat{a}^{\dag }(t_{1})\right\} \right\rangle ,
\end{eqnarray}%
In the case of dissipation, those Green's functions can be expressed as\cite%
{2015Diehl}%
\begin{eqnarray}
D^{R}\left( t,\omega \right)  &=&\frac{1}{\omega -\Delta -\Pi ^{R}\left(
t,\omega \right) +i\kappa },  \label{DR} \\
D^{A}\left( t,\omega \right)  &=&\frac{1}{\omega -\Delta -\Pi ^{A}\left(
t,\omega \right) -i\kappa },  \label{DA} \\
D^{K}\left( t,\omega \right)  &=&D^{R}\left( t,\omega \right) \left[ \Pi
^{K}\left( t,\omega \right) -2i\kappa \right] D^{A}\left( t,\omega \right) ,
\label{DK}
\end{eqnarray}%
Where $\Pi ^{K\left( R,A\right) }$ is the self-energy of the cavity field,
which can be calculated by%
\begin{eqnarray}
\Pi ^{R}\left( t_{2},t_{1}\right)  &=&-i\frac{1}{2}\lambda
^{2}\sum_{j_{1}j_{2}}\left[ G_{j_{1}+1,j_{2}+1}^{K}\left( t_{1},t_{2}\right)
G_{j_{2}j_{1}}^{R}\left( t_{2},t_{1}\right) +G_{j_{1}+1,j_{2}+1}^{A}\left(
t_{1},t_{2}\right) G_{j_{2}j_{1}}^{K}\left( t_{2},t_{1}\right) \right] ,
\label{se_ca_R} \\
\Pi ^{A}\left( t_{2},t_{1}\right)  &=&-i\frac{1}{2}\lambda
^{2}\sum_{j_{1}j_{2}}\left[ G_{j_{1}+1,j_{2}+1}^{R}\left( t_{1},t_{2}\right)
G_{j_{2}j_{1}}^{K}\left( t_{2},t_{1}\right) +G_{j_{1}+1,j_{2}+1}^{K}\left(
t_{1},t_{2}\right) G_{j_{2}j_{1}}^{A}\left( t_{2},t_{1}\right) \right] ,
\label{se_ca_A} \\
\Pi ^{K}\left( t_{2},t_{1}\right)  &=&-i\frac{1}{2}\lambda
^{2}\sum_{j_{1}j_{2}}\left[ G_{j_{1}+1,j_{2}+1}^{R}\left( t_{1},t_{2}\right)
G_{j_{2}j_{1}}^{A}\left( t_{2},t_{1}\right) +G_{j_{1}+1,j_{2}+1}^{A}\left(
t_{1},t_{2}\right) G_{j_{2}j_{1}}^{R}\left( t_{2},t_{1}\right) \right.
\notag \\
&&\left. +G_{j_{1}+1,j_{2}+1}^{K}\left( t_{1},t_{2}\right)
G_{j_{2}j_{1}}^{K}\left( t_{2},t_{1}\right) \right] .  \label{se_ca_K}
\end{eqnarray}%
We can see here for a free cavity, $\Pi ^{K\left( R,A\right) }=0$, and $%
D_{0}^{R(A)}\left( \omega \right) =\frac{1}{\omega -\Delta \pm i\kappa }$.
We substitute Eq.(\ref{se_ca_R})(\ref{se_ca_A})(\ref{se_ca_K}) and (\ref{DR}%
)(\ref{DA})(\ref{DK}) into Eq.(\ref{se_f_r})(\ref{se_f_a})(\ref{se_f_k}),
and make the Wigner transformation to obtain
\begin{eqnarray}
\Gamma _{j_{2},j_{1}}\left( t,\omega \right)  &=&i\left[ \Sigma
_{j_{2},j_{1}}^{R}\left( t,\omega \right) -\Sigma _{j_{2},j_{1}}^{A}\left(
t,\omega \right) \right]   \notag \\
&=&-\frac{1}{2}\lambda ^{2}\int \frac{d\nu }{2\pi }\left\vert D^{R}\left(
t,\omega \right) \right\vert ^{2}  \notag \\
&&\times \left\{ \left[ \Pi ^{R}\left( t,\nu \right) -\Pi ^{A}\left( t,\nu
\right) -2i\kappa \right] \left[ G_{j_{2}+1,j_{1}+1}^{K}\left( t,\omega -\nu
\right) -G_{j_{2}-1,j_{1}-1}^{K}\left( t,\omega +\nu \right) \right] \right.
\notag \\
&&\left. -i\left[ \Pi ^{K}\left( t,\nu \right) -2i\kappa \right] \left[
A_{j_{2}+1,j_{1}+1}\left( t,\omega -\nu \right) +A_{j_{2}-1,j_{1}-1}\left(
t,\omega +\nu \right) \right] \right\} ,  \label{life_time} \\
\Sigma _{j_{2},j_{1}}^{K}\left( t,\omega \right)  &=&i\frac{1}{2}\lambda
^{2}\int \frac{d\nu }{2\pi }\left\vert D^{R}\left( t,\omega \right)
\right\vert ^{2}  \notag \\
&&\times \left\{ i\left[ \Pi ^{R}\left( t,\nu \right) -\Pi ^{A}\left( t,\nu
\right) -2i\kappa \right] \left[ A_{j_{2}-1,j_{1}-1}\left( t,\omega +\nu
\right) -A_{j_{2}+1,j_{1}+1}\left( t,\omega -\nu \right) \right] \right.
\notag \\
&&\left. +\left[ \Pi ^{K}\left( t,\nu \right) -2i\kappa \right] \left[
G_{j_{2}+1,j_{1}+1}^{K}\left( t,\omega -\nu \right)
+G_{j_{2}-1,j_{1}-1}^{K}\left( t,\omega +\nu \right) \right] \right\} .
\label{self_energy}
\end{eqnarray}%
In the large dissipation limit, we approximate%
\begin{equation}
\left\vert D^{R}\left( t,\omega \right) \right\vert ^{2}\approx \left\vert
D_{0}^{R}\left( 0\right) \right\vert ^{2}=\frac{1}{\Delta ^{2}+\kappa ^{2}},
\end{equation}%
Then we substitute Eq.(\ref{life_time}) and (\ref{self_energy}) into Eq.(\ref%
{QKE_GK}). Keeping terms to order $\lambda ^{2}$, and integrating over $%
\omega $, we obtain the quantum kinetic equation for the single-particle
density matrix:
\begin{eqnarray}
\partial _{t}\rho _{ij}(t) &=&-i\lambda \left( \alpha ^{\ast }\rho
_{i+1,j-1}+\alpha \rho _{i-1,j}-\alpha ^{\ast }\rho _{i,j-1}-\alpha \rho
_{i,j+1}\right)   \notag \\
&&+\frac{2\kappa \lambda ^{2}}{\Delta _{c}^{\prime 2}+\kappa ^{2}}\left(
\rho _{i-1,j-1}-\rho _{i,j}\right)   \notag \\
&&+\frac{\kappa \lambda ^{2}}{\Delta _{c}^{\prime 2}+\kappa ^{2}}%
\sum_{l}\left( \rho _{i+1,l+1}\rho _{l,j}+\rho _{il}\rho _{l+1,j+1}-\rho
_{i-1,l-1}\rho _{\ell ,j}+\rho _{il}\rho _{l-1,j-1}\right) ,
\end{eqnarray}%
By solving this equation with Eq.(\ref{QKE_cavity}), we can obtain the
non-equilibrium dynamics including fluctuation effects.

\end{widetext}

\end{document}